\documentclass[conference]{IEEEtran}

\IEEEoverridecommandlockouts

\usepackage{cite}
\usepackage{bm}
\usepackage{amsmath,amssymb,amsfonts}
\usepackage{amsthm}        
\usepackage{algorithm}
\usepackage{algorithmic}
\usepackage{graphicx}
\usepackage{textcomp}
\usepackage[dvipsnames]{xcolor}
\usepackage{enumitem}
\usepackage{dblfloatfix}
\usepackage{float}
\usepackage{placeins}
\usepackage{adjustbox}
\usepackage{booktabs}
\usepackage{tabularx}
\usepackage{array}
\usepackage{multirow}
\usepackage{makecell}
\usepackage{tikz}
\usetikzlibrary{shapes,arrows}
\usepackage{mathtools}
\usepackage[normalem]{ulem}
\usepackage{subcaption}
\usepackage{caption}
\makeatletter
\let\paper@original@fs@ruled\fs@ruled
\renewcommand{\fs@ruled}{%
    \paper@original@fs@ruled
    \def\@fs@pre{\vspace{4pt}\hrule height.8pt depth0pt\kern2pt}%
    \def\@fs@mid{\kern2pt\hrule\kern1pt}%
    \def\@fs@post{\kern2pt\hrule\relax\vspace{-6pt}}%
}
\makeatother

\usepackage{siunitx}

\makeatletter
\renewcommand{\subsection}{\@startsection{subsection}{2}{\z@}%
    {1.0ex plus 0.5ex minus 0.2ex}%
    {0.45ex plus 0.2ex minus 0.1ex}%
    {\normalfont\normalsize\itshape}}
\makeatother

\usepackage[T1]{fontenc}
\usepackage{newtxtext,newtxmath}
\usepackage{microtype}

\usepackage{bbm} 
\usepackage[hidelinks,bookmarks=false]{hyperref} 
\usepackage{tikz}
\usetikzlibrary{arrows.meta,calc,positioning}
\usetikzlibrary{calc}
\newcolumntype{K}{>{\raggedright\arraybackslash}m{2.2 cm}}
\newcolumntype{E}{>{\centering\arraybackslash}m{2.3 cm}}
\newcolumntype{Y}{>{\centering\arraybackslash}m{4.2 cm}}
\newcolumntype{C}{>{\centering\arraybackslash}m{3.1 cm}}
\newcolumntype{d}{>{\centering\arraybackslash}m{1.35 cm}}
\newcolumntype{a}{>{\centering\arraybackslash}m{1.2 cm}}
\newcolumntype{e}{>{\centering\arraybackslash}m{1.35 cm}}
\newcolumntype{D}{>{\centering\arraybackslash}m{3 cm}}
\newcolumntype{z}{>{\centering\arraybackslash}m{0.7 cm}}
\newcolumntype{L}{>{\centering\arraybackslash}m{1 cm}}
\newcolumntype{x}{>{\centering\arraybackslash}m{0.75 cm}}
\newcolumntype{P}{>{\centering\arraybackslash}m{1.5 cm}}

\def\BibTeX{{\rm B\kern-.05em{\sc i\kern-.025em b}\kern-.08em
    T\kern-.1667em\lower.7ex\hbox{E}\kern-.125emX}}

\def\beq{\begin{equation}}
\def\eeq{\end{equation}}
\def\beqa{\begin{align}}
\def\eeqa{\end{align}}
\def\beqan{\begin{align*}}
\def\eeqan{\end{align*}}

\newtheorem{theorem}{Theorem}

\def\argmax{\mathop{\mathrm{arg\,max}}}

\def\tm1{t\! - \! 1}
\def\tp1{t\! + \! 1}

\AddToHookNext{shipout/foreground}{%
    \begin{tikzpicture}[remember picture,overlay]
        \node[anchor=north,align=center,font=\footnotesize,
              text width=\textwidth,inner xsep=0pt,inner ysep=4pt]
            at ([yshift=-0.3in]current page.north)
            {This work has been submitted to IEEE ICC 2027 for possible publication.};
    \end{tikzpicture}%
}

\begin{document}
\microtypesetup{activate=false}

\title{Intra-Cell Spatial Reuse in LEO Downlinks: Beam-Leakage-Aware Satellite Scheduling}

\author{
\IEEEauthorblockN{
Shizhen Jia\IEEEauthorrefmark{1}\IEEEauthorrefmark{2},
Hao Guo\IEEEauthorrefmark{2},
Ruoyu Sun\IEEEauthorrefmark{2},
Hans Geerdes\IEEEauthorrefmark{2},\\
Mark Poletti\IEEEauthorrefmark{2},
Songwei Dong\IEEEauthorrefmark{2},
Lauryn Smith\IEEEauthorrefmark{2}, and
Sundeep Rangan\IEEEauthorrefmark{1}
}

\IEEEauthorblockA{
\IEEEauthorrefmark{1}NYU WIRELESS, New York University, Brooklyn, NY, USA\\
\IEEEauthorrefmark{2}Department of Wireless Technologies, CableLabs, Louisville, Colorado, USA\\
}

\IEEEauthorblockA{\footnotesize
\{s.jia, srangan\}@nyu.edu, 
\{h.guo, r.sun, h.geerdes, m.poletti, songwei.dong, lauryn.smith\}@cablelabs.com
}
}

\bstctlcite{IEEEexample:BSTcontrol}

\maketitle
\begin{abstract}
Dense low-earth-orbit (LEO) constellations offer an opportunity to
increase local broadband capacity by serving multiple terminals in the
same cell through concurrent satellite beams. However, interference
between overlapping beams limits spatial reuse, and satellite selection
based solely on link strength or angular separation may result in suboptimal
performance. This paper proposes
beam-leakage-aware satellite scheduling for Ku-band LEO downlink transmissions.
With matched transmit and receive beamforming,
we formulate a fixed-cardinality sum-rate maximization problem that
jointly accounts for satellite-dependent received powers and pairwise
receive-array leakage. A beam-leakage-aware branch-and-bound (BL-BnB)
algorithm combines an admissible upper bound with greedy-plus-swap
initialization to obtain globally optimal schedules for the
reference-point model. We evaluate the method using a 24-h
ephemeris-derived Starlink visibility trace with antenna and transmit-power
models calibrated to public SpaceX filings. The results show that total
capacity increases with beam count, but less than proportionally.
With eight concurrent beams,
BL-BnB improves mean per-beam capacity by 27.2\% over nearest-satellite
selection, primarily benefiting weaker links. BL-BnB achieves
millisecond-scale runtimes in our evaluation, supporting the
computational feasibility of optimal scheduling.
\end{abstract}


\section{Introduction}
\label{sec:introduction}

\begin{figure}[!t]
    \vspace{-2pt}
    \centering
    \includegraphics[width=0.8\columnwidth]{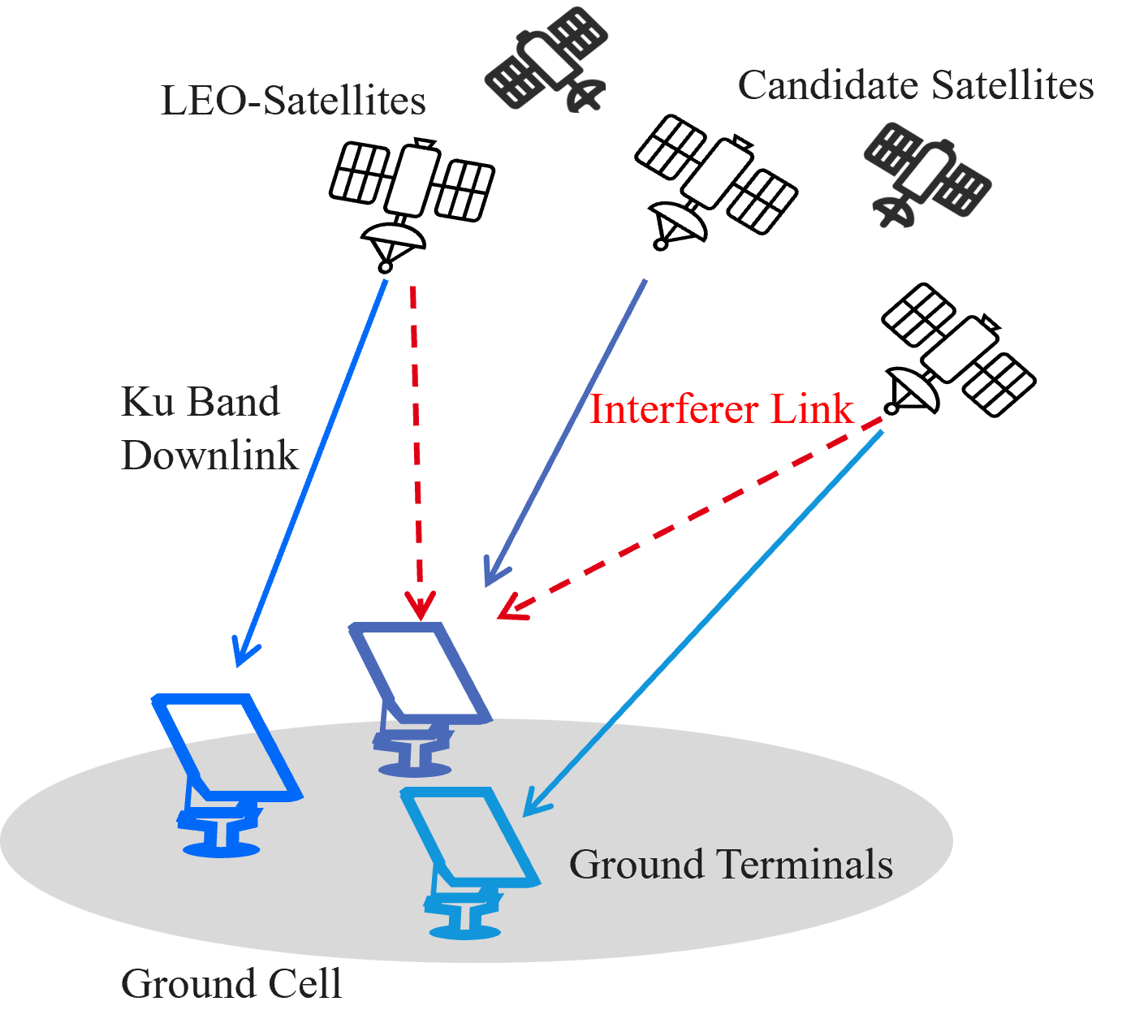}
    \par\vspace{-4pt}
    \caption{Considered Ku-band LEO downlink with multiple ground terminals
    and candidate satellites. Reuse of the same time--frequency resource
    creates inter-beam interference between concurrently active links.}
    \label{fig:system_overview}
    \vspace{-8pt}
\end{figure}

Local traffic demand in low-Earth-orbit (LEO) broadband networks can exceed the capacity
of a single satellite beam, motivating concurrent service from multiple
simultaneously visible satellites~\cite{kim2025ephemeris}. In a recreational
vehicle (RV) park or at rural terminal densities, active terminals can be
too close together to be served by non-overlapping spot beams.
Concurrent beams from multiple satellites can increase local capacity
by reusing the same time--frequency resource. The resulting intra-cell
spatial reuse, illustrated in
Fig.~\ref{fig:system_overview}, depends on how effectively the terminals
can separate desired and interfering downlinks.

Regulatory developments further motivate the study of overlapping
beams. To protect geostationary-satellite-orbit (GSO) systems under the
International Telecommunication Union (ITU)'s interference
limits~\cite{itu2024radio}, the Federal Communications Commission (FCC)
initially restricted Starlink Gen2 to one co-frequency, co-polarization
beam per location in the 12.2--12.7 and 19.7--20.2~GHz
bands~\cite{fcc2022starlinkgen2}. Then, its January 2026 waiver permitted up to
eight overlapping Ku-band beams with $4^\circ$ GSO-arc
avoidance~\cite{fcc2026starlinkwaiver}. The framework adopted in April
2026 introduced an alternative that does not explicitly limit the number
of overlapping beams but instead relies on GSO-link performance
protection and $3^\circ$ arc avoidance in the United
States~\cite{fcc2026spectrumsharing}. These developments make the
capacity benefits and interference costs of concurrent beams
increasingly relevant.

The most critical challenge is that satellite visibility does not directly
translate into usable spatial reuse. For a compact cell, satellite
transmit footprints substantially overlap, so interference rejection
relies primarily on the ground terminals' receive arrays. Selecting the
nearest satellites favors short propagation paths but can also end up with
highly correlated arrival directions. Conversely, enforcing a large
angular separation can be unnecessarily restrictive, sacrificing
desired-link strength without minimizing the actual array response to
interference. Since each active
satellite contributes interference to all other scheduled links, the
achievable sum rate depends jointly on individual received powers and
aggregate beam leakage.

Prior work has explored spatial processing and resource allocation to
control satellite interference. Massive multiple-input multiple-output
(MIMO) transmission combines statistical channel information and
space-angle-based user grouping for full-frequency reuse, but considers
a single satellite serving all scheduled users~\cite{you2020massive}.
Cooperative downlinks jointly optimize hybrid beamforming and user
scheduling, assuming inter-satellite data exchange~\cite{zhang2024cooperative}.
Game-theoretic transmit covariance optimization mitigates uplink
jamming, assuming cooperative satellite reception~\cite{NTN-Antijamming}.
Interference prediction and hybrid receive beamforming improve long-term
rates through receiver adaptation rather than satellite subset
selection~\cite{xin2025interference}. Coordinated beam hopping balances
loads using spatially isolated transmission patterns~\cite{lin2023beamhopping},
restricting simultaneous illumination of the same cell.
These approaches leave open the achievable intra-cell reuse from
satellite selection alone with fixed matched beamformers.

More closely related, ephemeris-based min--max satellite selection
(E-MMS) greedily minimizes the largest receive-array correlation with
already selected satellites~\cite{kim2025ephemeris}.
This criterion measures spatial separability. However, the desired
received power (the beamformed signal power from the intended
satellite) varies across candidate links with propagation distance,
antenna gain, and transmit power. Each link's rate also depends on the
sum of interference from all other selected satellites, which the
largest pairwise correlation alone does not capture.
Thus, correlation-based selection does not guarantee the maximum sum
rate. Computing the globally optimal sum rate under the same model
provides a reference for assessing heuristic selection.

This paper studies concurrent Ku-band LEO downlinks with matched
transmit and receive beamforming. We select a prescribed number of
satellites at each scheduling interval using a common cell reference
point. A gateway can estimate the scheduling inputs from satellite
ephemerides, terminal Global Positioning System (GPS) coordinates,
known beam powers and array responses, and precomputed loss tables
based on ITU Radiocommunication Sector (ITU-R) Recommendation
P.618~\cite{iturp61814}. Our contributions are threefold:
\begin{itemize}[leftmargin=*,itemsep=2pt,topsep=3pt]
    \item \textbf{Beam-leakage-aware scheduling formulation:}\\
    We formulate fixed-cardinality intra-cell sum-rate maximization
    using satellite-dependent received powers and pairwise receive-array
    leakage under full time--frequency reuse.

    \item \textbf{Globally optimal satellite selection:}\\
    Beam-leakage-aware branch-and-bound (BL-BnB) combines an admissible upper bound with greedy-plus-swap
    initialization to certify optimality for the reference-point model.
    Selecting eight of 28 candidates is approximately $267\times$
    faster than exhaustive search on average.

    \item \textbf{Practical spatial-reuse assessment:}\\
    A 24-h Starlink trace and models calibrated to public SpaceX
    filings quantify the effects of beam count, cell radius, and
    scheduling. Mean per-beam capacity improves over nearest-satellite
    selection by 27.2\% with eight beams and 9.8\% with 49 beams,
    primarily benefiting weaker links.
\end{itemize}

\section{Problem Formulation}
\label{sec:problem_formulation}

We consider an LEO downlink where multiple visible satellites serve a
ground cell on a shared time–frequency resource. In each scheduling
interval, geometry and large-scale channel states are fixed, and $K$
concurrent layers are selected to maximize the instantaneous cell sum
rate, accounting for link strength and inter-layer beam leakage.

\subsection{System Model and Assumptions}

At time instant $t$, let $\mathcal S(t)$ denote the set of satellites
that satisfy the visibility and link-availability requirements, and denote
$N(t)=\left|\mathcal S(t)\right|$ as the number of satellites in this set. We focus on intervals with $N(t)\ge K$, and if fewer
than $K$ satellites are visible, the scheduler activates all available
satellites and replaces $K$ by $N(t)$. Each selected satellite provides
one downlink layer to the cell. Because the cell diameter is much smaller
than the LEO slant range, the scheduling metrics are evaluated at a
common cell reference point $\mathbf r_{\mathrm c}$, assuming identical
receive arrays with the same physical orientation. Each satellite carries independent data
for one terminal and satellite--user assignment is fixed.

For satellite $s$ at point $\mathbf{r_s}$, the slant range is
$d_s(t)=\|\mathbf r_s(t)-\mathbf r_{\mathrm c}\|_2$, giving free-space
loss $\beta_s^{\mathrm{FSPL}}(t)=(\lambda/(4\pi d_s(t)))^2$ at
wavelength $\lambda$. The effective large-scale gain is
\begin{equation}
\beta_s^{\mathrm{eff}}(t)
=
\beta_s^{\mathrm{FSPL}}(t)
A_s^{\mathrm{atm}}(t)
A_s^{\mathrm{rain}}(t)
A_s^{\mathrm{pol}}(t)
A_s^{\mathrm{misc}}(t),
\label{eq:effective_gain}
\end{equation}
where the factors account for non-rain atmospheric, rain,
polarization, and other large-scale losses.

Each satellite employs an $N_{\text{t}}$-element transmit array, and each ground
receiver employs an $N_{\text{r}}$-element receive array. Under the dominant LoS
assumption, the channel associated with satellite $s$ is
\begin{equation}
\mathbf H_s(t)
=
\sqrt{\beta_s^{\mathrm{eff}}(t)G_{\text{t},s}(t)G_{\text{r},s}(t)}
\mathbf a_{\text{r},s}(t)\mathbf a_{\text{t},s}^{\mathrm H}(t),
\label{eq:los_channel}
\end{equation}
where $G_{\text{t},s}$ and $G_{\text{r},s}$ are linear element-pattern gains in the
departure and arrival directions, respectively. A common link phase is
omitted because it cancels in the received-power expressions.
The transmit and receive steering vectors, i.e., $\mathbf a_{\text{t},s}(t)$ and $\mathbf a_{\text{r},s}(t)$, have squared norms
$N_{\text{t}}$ and $N_{\text{r}}$, respectively.
For the rank-one channel in~\eqref{eq:los_channel}, dominant-mode
singular value decomposition (SVD) beamforming is equivalent to matched steering,
$\mathbf f_s=\mathbf a_{\text{t},s}/\sqrt{N_{\text{t}}}$ and
$\mathbf w_s=\mathbf a_{\text{r},s}/\sqrt{N_{\text{r}}}$. With transmit power
$P_{\mathrm{tx},s}$, the desired post-beamforming received power is therefore
\begin{equation}
q_s(t)=P_{\mathrm{tx},s}\beta_s^{\mathrm{eff}}(t)
G_{\text{t},s}(t)G_{\text{r},s}(t)N_{\text{t}}N_{\text{r}}.
\label{eq:desired_power}
\end{equation}

\subsection{Beamformed Signal and Interference Model}

When the receiver is steered toward satellite $s$, a cochannel signal
from satellite $s'$ is attenuated according to the receive-array
de-correlation. We define the normalized leakage coefficient as
\begin{equation}
\eta_{s,s'}(t)
=
\frac{\left|\mathbf a_{\text{r},s}^{\mathrm H}(t)
\mathbf a_{\text{r},s'}(t)\right|^2}{N_{\text{r}}^2},
\qquad 0\le\eta_{s,s'}(t)\le 1.
\label{eq:beam_leakage}
\end{equation}
Consequently, the interference power from satellite $s'$ to the layer
associated with satellite $s$ is $q_{s'}(t)\eta_{s,s'}(t)$.

For a scheduled subset $\mathcal A(t)\subseteq\mathcal S(t)$ with
$|\mathcal A(t)|=K$, the signal-to-interference-plus-noise ratio (SINR)
of layer $s$ is
\begin{equation}
\gamma_s(\mathcal A,t)
=
\frac{q_s(t)}{
\sigma^2+
\sum\limits_{\substack{s'\in\mathcal A(t)\\s'\neq s}}
q_{s'}(t)\eta_{s,s'}(t)}.
\label{eq:scheduling_sinr}
\end{equation}
Here, $\sigma^2=N_0B$, where $N_0$ is the effective noise power spectral
density, including the receiver noise figure, and $B$ is the bandwidth.
Treating inter-layer interference as noise,
the per-layer rate is
$R_s(\mathcal A,t)=B\log_2(1+\gamma_s(\mathcal A,t))$,
and the cell sum rate is
$C_{\mathrm{sum}}(\mathcal A,t)=\sum_{s\in\mathcal A(t)}R_s(\mathcal A,t)$.

\subsection{Scheduling Objective}

The scheduler selects exactly $K$ satellites to maximize the cell
sum rate. Let $y_s(t)$ indicate whether satellite $s$ is scheduled,
so that $\mathcal A(t)=\left\{s\in\mathcal S(t):y_s(t)=1\right\}$. Using the SINR
in~\eqref{eq:scheduling_sinr}, we solve
\begin{subequations}
\label{prob:satellite_scheduling}
\begin{align}
\mathbf y^\star(t)
&\in
\argmax_{\mathbf y(t)}
\sum_{s\in\mathcal S(t)}y_s(t)B
\log_2\!\bigl(1+\gamma_s(\mathcal A,t)\bigr)
\label{eq:binary_scheduling_objective}
\\
\mathrm{s.t.}\quad
&\sum_{s\in\mathcal S(t)}y_s(t)=K,\quad
 y_s(t)\in\{0,1\},\ \forall s\in\mathcal S(t).
\label{eq:binary_scheduling_constraints}
\end{align}
\end{subequations}
The objective is nonseparable because each selected satellite changes
the interference at the other scheduled layers. Maximizing the sum
rate therefore requires jointly accounting for the received powers
$\{q_s(t)\}$ and pairwise leakage coefficients $\{\eta_{s,s'}(t)\}$.

\section{Optimal Scheduling}
\label{sec:optimal_scheduling}

Problem~\eqref{prob:satellite_scheduling} admits
$\binom{N(t)}{K}$ candidate subsets. We obtain the globally optimal
schedule using BL-BnB (Algorithm~\ref{alg:bl_bnb}).
It maintains the best feasible schedule found so far (the incumbent)
and prunes branches whose admissible upper bounds do not exceed its
sum rate.

\begin{algorithm}[!t]
\caption{BL-BnB Satellite Scheduling}
\label{alg:bl_bnb}
\footnotesize
\begin{algorithmic}[1]
\REQUIRE $\mathcal S$, $\{q_s\}$, $\{\eta_{s,j}\}$, $K$, $B$, and
$\sigma^2$
\ENSURE Globally optimal set $\mathcal A^\star$ and sum rate $C^\star$
\STATE Set $g_{s,j}\gets q_j\eta_{s,j}$ and $g_{s,s}\gets0$
\STATE Initialize $\mathcal A^\star$ by maximizing $C_{\mathrm{sum}}$
      over the top-$K$ and bottom-$K$ power sets and the greedy--swap set
\STATE $C^\star\gets C_{\mathrm{sum}}(\mathcal A^\star)$ and
      $\mathcal Q\gets[(\emptyset,\mathcal S)]$
\WHILE{$\mathcal Q\neq\emptyset$}
    \STATE Pop $(\mathcal P,\mathcal R)$ from $\mathcal Q$ and set
          $m\gets K-|\mathcal P|$
    \IF{$m<0$ or $m>|\mathcal R|$}
        \STATE \textbf{continue}
    \ELSIF{$m=0$}
        \STATE $\mathcal A\gets\mathcal P$
    \ELSIF{$m=|\mathcal R|$}
        \STATE $\mathcal A\gets\mathcal P\cup\mathcal R$
    \ELSE
        \STATE Compute $\mathrm{UB}(\mathcal P,\mathcal R)$ using
              \eqref{eq:bnb_upper_bound}
        \IF{$\mathrm{UB}(\mathcal P,\mathcal R)\le C^\star$}
            \STATE \textbf{continue}
        \ENDIF
        \STATE $c^\star\gets\argmax\limits_{c\in\mathcal R}
              \bar R_c^{(\mathcal P,\mathcal R)}$ from \eqref{eq:candidate_rate_bound}
        \STATE $\mathcal R'\gets\mathcal R\setminus\{c^\star\}$
        \STATE Push $(\mathcal P,\mathcal R')$ and then
              $(\mathcal P\cup\{c^\star\},\mathcal R')$ onto $\mathcal Q$
        \STATE \textbf{continue}
    \ENDIF
    \STATE $C\gets C_{\mathrm{sum}}(\mathcal A)$
    \STATE If $C>C^\star$, set $(\mathcal A^\star,C^\star)\gets(\mathcal A,C)$
\ENDWHILE
\RETURN $\mathcal A^\star$ and $C^\star$
\end{algorithmic}
\end{algorithm}

\subsection{Beam-Leakage-Aware Upper Bound}

The time index is omitted in this section. A search node
$(\mathcal P,\mathcal R)$ consists of a partial selection $\mathcal P$
and a disjoint pool of undecided candidates $\mathcal R$.
With $m=K-|\mathcal P|$, the node is infeasible if $m<0$ or $m>|\mathcal R|$.
Each choice of $\mathcal T\subseteq\mathcal R$ with $|\mathcal T|=m$
completes the schedule as $\mathcal A=\mathcal P\cup\mathcal T$,
where $|\mathcal A|=K$.

For a satellite $s$ and candidate set $\mathcal U$, define the minimum
additional interference contributed by any $m$ members of $\mathcal U$
as
\begin{equation}
L_s(\mathcal U,m)
=
\min_{\substack{\mathcal B\subseteq\mathcal U\\|\mathcal B|=m}}
\sum_{j\in\mathcal B}q_j\eta_{s,j},
\qquad L_s(\mathcal U,0)=0.
\label{eq:minimum_interference}
\end{equation}
Thus, $L_s(\mathcal U,m)$ is the sum of the $m$ smallest values in
$\{q_j\eta_{s,j}:j\in\mathcal U\}$.

For an already selected satellite $s\in\mathcal P$, define an upper
bound on its rate in any feasible completion as
\begin{equation}
\bar R_s^{(\mathcal P,\mathcal R)}
=
B\log_2\!\left(
1+
\frac{q_s}{
\sigma^2+
\sum\limits_{\substack{j\in\mathcal P\\j\neq s}}
q_j\eta_{s,j}
+L_s(\mathcal R,m)}
\right).
\label{eq:selected_rate_bound}
\end{equation}
For $m\ge1$ and an undecided satellite $c\in\mathcal R$, selection of $c$ requires
$m-1$ additional satellites from $\mathcal R\setminus\{c\}$, giving
\begin{equation}
\bar R_c^{(\mathcal P,\mathcal R)}
=
B\log_2\!\left(
1+
\frac{q_c}{
\sigma^2+
\sum\limits_{j\in\mathcal P}q_j\eta_{c,j}
+L_c(\mathcal R\!\setminus\!\{c\},m-1)}
\right).
\label{eq:candidate_rate_bound}
\end{equation}
Let $\mathrm{Top}_m(\mathcal R)$ contain the $m$ undecided satellites
with the largest values of~\eqref{eq:candidate_rate_bound}. An upper
bound on the best sum rate reachable from node
$(\mathcal P,\mathcal R)$ with $m\ge1$ is
\begin{equation}
\mathrm{UB}(\mathcal P,\mathcal R)
=
\sum_{s\in\mathcal P}\bar R_s^{(\mathcal P,\mathcal R)}
+
\sum_{c\in\mathrm{Top}_m(\mathcal R)}
\bar R_c^{(\mathcal P,\mathcal R)}
\label{eq:bnb_upper_bound}
\end{equation}

\begin{theorem}
For every feasible completion $\mathcal T\subseteq\mathcal R$ of a
nonterminal node with $|\mathcal T|=m\ge1$,
\begin{equation}
C_{\mathrm{sum}}(\mathcal P\cup\mathcal T)
\le
\mathrm{UB}(\mathcal P,\mathcal R)
\end{equation}
\end{theorem}

\paragraph*{Proof}
For each $s\in\mathcal P$, the interference contributed by any feasible
completion $\mathcal T$ is no smaller than
$L_s(\mathcal R,m)$. Similarly, for each $c\in\mathcal T$, the
interference from the other $m-1$ newly selected satellites is no
smaller than $L_c(\mathcal R\setminus\{c\},m-1)$. Because the rate is
monotonically decreasing in interference, the corresponding rates are
bounded by~\eqref{eq:selected_rate_bound} and
\eqref{eq:candidate_rate_bound}. Finally, any completion contains
exactly $m$ candidates, whose total contribution cannot exceed the sum
of the $m$ largest candidate bounds in~\eqref{eq:bnb_upper_bound}.

\subsection{Branch-and-Bound Search and Initialization}

The search starts at $(\emptyset,\mathcal S)$. A node is
infeasible if $m<0$ or $m>|\mathcal R|$, meaning that too many
satellites have been selected or too few candidates remain to
reach $K$. Such nodes are discarded. If $m=0$, the only completion
is $\mathcal P$; if $m=|\mathcal R|$, it is
$\mathcal P\cup\mathcal R$. In either case, its sum rate is
evaluated using the original objective, and the incumbent is
updated if improved.

For a nonterminal node, pruning occurs when
$\mathrm{UB}(\mathcal P,\mathcal R)\le C^\star$, where $C^\star$
is the best feasible sum rate found so far. By the preceding
theorem, no completion of this node can improve $C^\star$.
Otherwise, the candidate $c^\star$ specified in
Algorithm~\ref{alg:bl_bnb} generates two children:
$(\mathcal P\cup\{c^\star\},\mathcal R\setminus\{c^\star\})$
and $(\mathcal P,\mathcal R\setminus\{c^\star\})$.
The inclusion child is explored first to seek early incumbent
improvements. Once the search is exhausted, the incumbent is
globally optimal.

Initialization greedily adds the candidate yielding the largest
resulting sum rate until $K$ satellites are selected. One-swap
refinement then repeatedly accepts the best strictly improving
exchange between a selected and an unselected satellite until none
remain. The best of this set and the top-$K$ and bottom-$K$
received-power sets initialize the incumbent. This affects search
efficiency, not optimality. Algorithm~\ref{alg:bl_bnb} specifies the
branching rule and complete procedure.

\subsection{Complexity Analysis}

Precompute
$g_{s,s'}=q_{s'}\eta_{s,s'}$. In this way, evaluating one subset of size $K$ 
requires $\mathcal O\left(K^2\right)$ operations, giving exhaustive cost
$\mathcal O\left(\binom{N}{K}K^2\right)$ and interference-matrix storage
$\mathcal O\left(N^2\right)$.

Computing all pairwise receive-array correlations by explicit inner
products of length $N_{\text{r}}$ requires
$\mathcal O\left(N^2N_{\text{r}}\right)$ operations. At each BL-BnB
node, the implemented upper-bound calculation fully sorts up to $N$
interference lists, each containing at most $N$ values. Its worst-case
cost is therefore $\mathcal O\left(N^2\log N\right)$.
If $M_{\mathrm{BB}}$ nodes are visited and
$I_{\mathrm{sw}}$ swap iterations are performed, the overall
complexity is
\begin{equation}
\mathcal O\!\left(
N^2N_{\text{r}}
+(1+I_{\mathrm{sw}})NK^3
+M_{\mathrm{BB}}N^2\log N
\right)
\end{equation}
BL-BnB remains exponential in the worst case as $M_{\mathrm{BB}}$
may approach exhaustive enumeration, while admissible pruning reduces
practical search cost while preserving global optimality.

\begin{figure}[!htbp]
    \vspace{-2pt}
    \centering
    \includegraphics[width=\columnwidth]{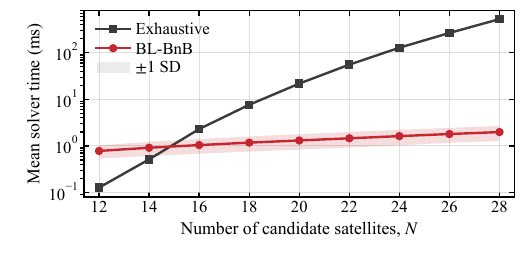}
    \par\vspace{-4pt}
    \caption{Solver runtime (ms) in log-scale for selecting $K=8$ of $N$ candidates.
    Curves show the mean of 720 per-snapshot mean runtimes (three
    repeats per method and snapshot). Shading indicates $\pm1$ sample
    standard deviation across these means.}
    \label{fig:bnb_runtime_scaling}
    \vspace{-8pt}
\end{figure}
Fig.~\ref{fig:bnb_runtime_scaling} compares BL-BnB and vectorized
exhaustive search on identical strongest-$N$ sets ($K=8$), timing only
initialization and search. Initialization and bounding overhead make
BL-BnB slower at small $N$, but it is faster at every tested $N\ge16$.
At $N=28$, mean runtime falls from 529.48 to 1.98~ms (about
$267\times$). With all 49--90 visible candidates, BL-BnB takes
4.83~ms (median of 720 per-snapshot medians, each from three repeats).

\section{Practical Study and Performance Analysis}
\label{sec:simulation}

\subsection{Simulation Methodology}
We evaluate 720 snapshots sampled at 2-min intervals over a 24-h
Starlink trace generated from CelesTrak orbital
elements~\cite{celestrakstarlink} (Table~\ref{tab:simulation_settings}),
with 49--90 visible satellites per snapshot.
Both methods select $K$ visible satellites, with one beam per satellite
serving an independently generated user on the shared time--frequency
resource. The baseline chooses the
$K$ nearest satellites, whereas BL-BnB maximizes the reference-point
sum rate. Using the same SINR model, we apply the 3GPP
implementation-loss factor of 0.6~\cite{3gpp36942} and set the SINR
cap to 30 dB. The objective, initialization, bounds, and evaluation
use the resulting nondecreasing mapping
$r(\gamma)=0.6B\log_2\left(1+\min\left\{\gamma,10^3\right\}\right)$.
Evaluation uses user locations.
Unless noted, $R=250$ m.
The rates provide a snapshot-level benchmark for spatial reuse.
Handover and other unmodeled overhead would reduce net throughput.
For each elevation angle, ITU-R P.618 provides the attenuation level
exceeded for 1\% of an average year. These values are precomputed
on a $0.1^\circ$ elevation grid.

\begin{table}[!t]
    \vspace{3pt}
    \centering
    \caption{Simulation Parameters.}
    \label{tab:simulation_settings}
    \vspace{-3pt}
    \footnotesize
    \setlength{\tabcolsep}{3.5pt}
    \renewcommand{\arraystretch}{1.08}
    \begin{tabularx}{\columnwidth}{|>{\raggedright\arraybackslash}p{0.34\columnwidth}|>{\centering\arraybackslash}X|}
        \hline
        \textbf{Radio Parameters} & \textbf{Values} \\
        \hline
        Carrier/Bandwidth
            & \SI{11.7}{\giga\hertz}/\SI{250}{\mega\hertz} \\
        \hline
        Noise PSD/Noise Figure
            & $-174$ dBm/Hz/5 dB \\
        \hline
        TX/RX Arrays
            & $30\times30$/$20\times23$ uniform rectangular arrays (URAs) \\
        \hline
        Channel/Beamforming
            & Matched steering, ITU-R P.618 losses~\cite{iturp61814} \\
        \hline
        \multicolumn{2}{|l|}{\textbf{Scene Configuration}} \\
        \hline
        Reference Location
            & $39.95697^{\circ}$N, $105.16033^{\circ}$W \\
        \hline
        Duration/Sampling
            & 24 h, 720 snapshots, 2-min interval \\
        \hline
        Orbit/Elevation Mask
            & Gen-1/Gen-2 LEO, $25^{\circ}$ \\
        \hline
        Cell Radius
            & $R\in\{0.25,5,15\}$ km \\
        \hline
        Paired Beams
            & $K\in\{1,4,8,14,20,30,40,49\}$ \\
        \hline
        Spectrum Reuse
            & Full time--frequency reuse \\
        \hline
    \end{tabularx}
    \par\vspace{-4pt}
\end{table}

\begin{figure}[!htbp]
    \vspace{-2pt}
    \centering
    \includegraphics[width=\columnwidth]{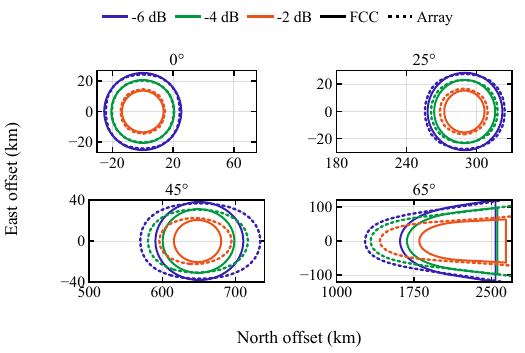}
    \par\vspace{-4pt}
    \caption{Ground gain contours of SpaceX Gen2~\cite{spacex2020gen2technical}
    (solid) and the modeled $30\times30$ array (dotted) at 11.7 GHz and
    614-km altitude. Scan angles: $0^{\circ}$, $25^{\circ}$, $45^{\circ}$,
    $65^{\circ}$. Contour levels: $-6$, $-4$, $-2$ dB (colored).}
    \label{fig:fcc_array_footprint}
    \vspace{-8pt}
\end{figure}

\begin{figure}[!htbp]
    \vspace{-2pt}
    \centering
    \includegraphics[width=0.98\columnwidth]{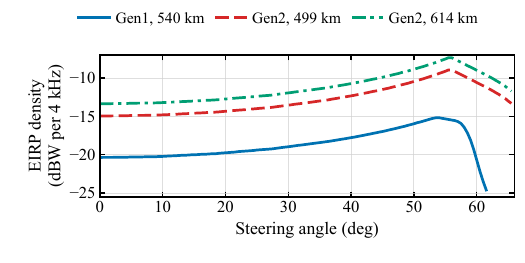}
    \par\vspace{-4pt}
    \caption{Starlink EIRP density versus steering angle for Gen-1 at
    540 km~\cite{spacex2020gen1technical} and Gen-2 at 499 km (approximated)
    and 614 km~\cite{spacex2020gen2technical}, in dBW per 4 kHz.}
    \label{fig:eirp_steering_angle}
    \vspace{-8pt}
\end{figure}
We model and calibrate the $30\times30$ array to SpaceX Gen2 low-gain Ku-band
beams~\cite{spacex2020gen2technical}, adjusting element gain for
34-dBi boresight gain under the conditions in
Fig.~\ref{fig:fcc_array_footprint}. Modeled contours closely match the filing
at $0^{\circ}$ and $25^{\circ}$ but extend beyond it at $45^{\circ}$
and $65^{\circ}$. Although larger arrays could narrow these footprints,
we retain the fixed size because the studied cells ($R\leq15$ km)
remain within the filed $-2$-dB contours at both angles.
For the RV-park terminals, the $20\times23$ array approximates Starlink
Mini's $298.5\times259$-mm dimensions~\cite{starlinkminispec} and matches
the 30.4-dBi receive gain of SpaceX's ESIM3-V2
terminal~\cite{fcc2024smallterminal}.

Transmit power follows the digitized steering-dependent equivalent
isotropically radiated power (EIRP) curves~\cite{spacex2018technical}
in Fig.~\ref{fig:eirp_steering_angle}. The 499-km curve shifts
the 614-km curve down by 1.6 dB to match the filed maximum EIRP density.

\begin{figure}[!t]
    \vspace{-2pt}
    \centering
    \includegraphics[width=0.98\columnwidth]{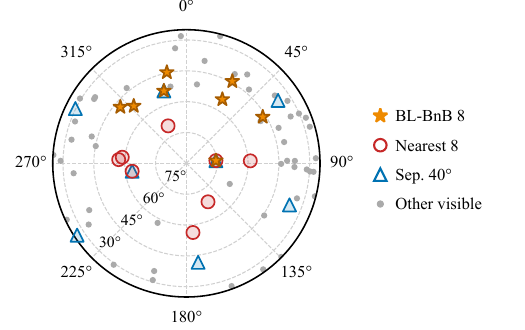}
    \par\vspace{-4pt}
    \caption{Satellite-selection sky view: overlaid markers indicate shared
    satellites. Gray dots denote other visible satellites. Outer labels
    and rings indicate azimuth and elevation, respectively.}
    \label{fig:satellite_selection_sky}
    \vspace{-8pt}
\end{figure}

\subsection{Intra-cell Performance}

\begin{figure}[!t]
    \vspace{-2pt}
    \centering
    \includegraphics[width=0.98\columnwidth]{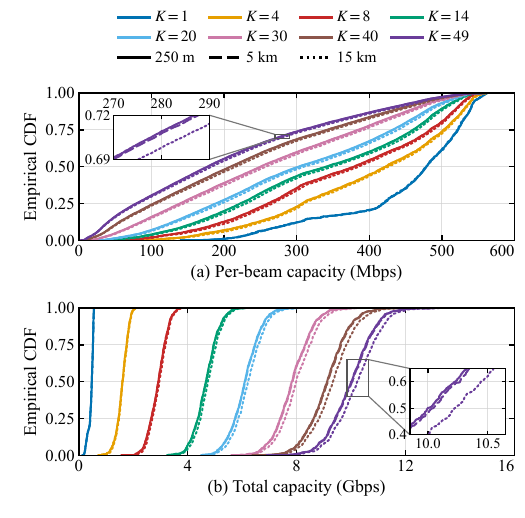}
    \par\vspace{-4pt}
    \caption{Empirical CDFs of (a) per-beam and (b) total capacity.
    Colors denote the number of paired beams $K$. Line styles denote
    cell radii of 0.25, 5, and 15 km. Insets magnify the boxed regions
    for $K=49$.}
    \label{fig:capacity_cdfs}
    \vspace{-6pt}
\end{figure}

In Fig.~\ref{fig:capacity_cdfs}, at $R=250$ m and $K=1,8,49$, mean
per-beam capacities are 451, 363, and 204 Mbps, while mean totals are
0.45, 2.90, and 9.98 Gbps, respectively. Thus, 49 overlapping beams
provide only $22.2\times$ the single-beam capacity, with declining
per-beam rates. Additional beams reuse the full bandwidth but
increase aggregate receive-array leakage, limiting spatial
multiplexing gains.

Increasing $R$ to 5 km changes the mean by less than
0.3\% for every $K$. As the insets show, differences emerge mainly
at 15 km with many beams, where the $K=49$ mean total rises from 9.98 to 10.21 Gbps
(2.3\%). The weak radius dependence is consistent with modest
variations in slant range and arrival geometry. With substantially
overlapping satellite main lobes,
interference rejection relies primarily on receive beamforming to
resolve arrival directions.

\subsection{Optimal Scheduling Performance}

\begin{figure}[!t]
    \vspace{-2pt}
    \centering
    \includegraphics[width=0.95\columnwidth]{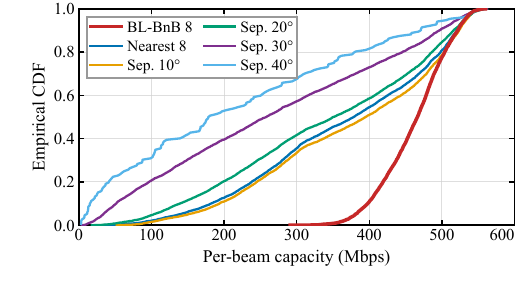}
    \par\vspace{-4pt}
    \caption{Per-beam capacity CDFs for eight paired beams under BL-BnB,
    nearest-satellite selection, and spherical-separation baselines.
    The $40^{\circ}$ curve uses only the 19 snapshots where eight satellites
    can be selected.}
    \label{fig:selection_policy_capacity_cdf}
    \vspace{-8pt}
\end{figure}

The eight-beam policy CDF uses matched inputs. Beam-count comparisons
include unpaired archived user scenes at $K=8,14,30$ and slightly
different propagation losses.

Nearest-eight selection favors high elevations, potentially clustering
arrival directions and increasing receive-array correlation
(Fig.~\ref{fig:satellite_selection_sky}). In
Fig.~\ref{fig:selection_policy_capacity_cdf}, a $10^{\circ}$ minimum
spherical separation raises mean per-beam capacity from 362.9 to
372.9 Mbps (2.8\%), whereas $20^{\circ}$ and $30^{\circ}$ reduce it
to 339.8 and 267.7 Mbps. Thus, wider separation alone does not optimize
the signal--interference tradeoff. E-MMS~\cite{kim2025ephemeris} instead
greedily minimizes maximum pairwise receive-array correlation,
differing from our fixed-angle baseline without guaranteeing sum-rate
optimality. BL-BnB optimizes reference-point sum capacity globally,
averaging 461.6 Mbps per beam (27.2\% above nearest-eight selection).

\begin{figure}[!htbp]
    \vspace{-2pt}
    \centering
    \includegraphics[width=0.98\columnwidth]{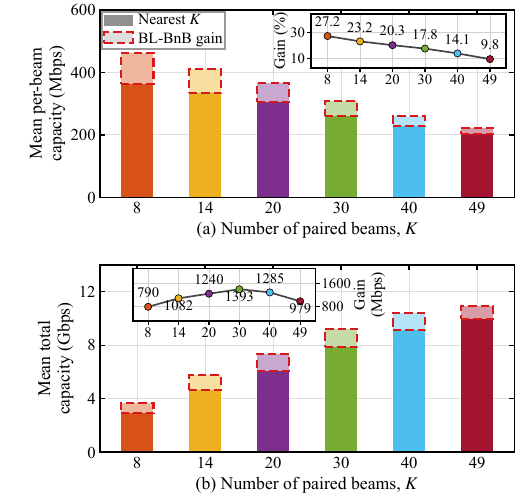}
    \par\vspace{-4pt}
    \caption{Mean (a) per-beam and (b) total capacity versus $K$.
    Colored bars show nearest-$K$ selection. Red dashed extensions show
    BL-BnB gains. Insets show (a) percentage and (b) absolute total-capacity
    gains over nearest-$K$ selection.}
    \label{fig:mean_capacity_tradeoff}
    \vspace{-6pt}
\end{figure}

In Fig.~\ref{fig:mean_capacity_tradeoff}(a), mean per-beam gain falls
from 98.7 Mbps at $K=8$ to 20.0 Mbps at $K=49$. Relative gains are
27.2\%, 17.8\%, and 9.8\% at $K=8,30,49$, respectively. Activating more visible
satellites leaves fewer choices to avoid weak links or strong
interference. Total gain equals $K$ times the mean per-beam gain:
790, 1393, and 979 Mbps at $K=8,30,49$, respectively
(see Fig.~\ref{fig:mean_capacity_tradeoff}(b)). Beyond $K=30$, declining
per-beam gains outweigh the increasing beam count.

\begin{figure}[!htbp]
    \vspace{-2pt}
    \centering
    \includegraphics[width=0.98\columnwidth]{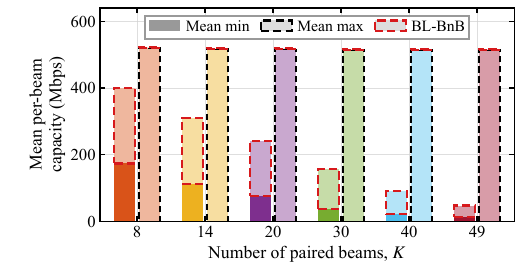}
    \par\vspace{-4pt}
    \caption{Time-averaged per-snapshot minimum (solid) and maximum
    (black dashed) beam capacities versus $K$ for nearest-$K$ selection.
    Red dashed outlines show BL-BnB values.}
    \label{fig:mean_min_max_per_beam_capacity}
    \vspace{-8pt}
\end{figure}

\FloatBarrier
For $K=8$, BL-BnB raises the time-averaged per-snapshot maximum
capacity from 520.6 to 523.8 Mbps (0.6\%) and the minimum from 173.8
to 401.3 Mbps (Fig.~\ref{fig:mean_min_max_per_beam_capacity}). This
suggests that better satellite choices and interference avoidance
mainly benefit weaker links. Consistent with
Fig.~\ref{fig:mean_capacity_tradeoff}, the minimum-capacity gain falls
from 227.5 to 34.4 Mbps as $K$ increases from 8 to 49.

\section{Conclusion}
\label{sec:conclusion}

We studied intra-cell spatial reuse in LEO downlinks and developed BL-BnB
to jointly account for link strength and receive-array leakage. The method
combines branch-and-bound with an admissible upper bound and greedy-plus-swap
initialization to obtain globally optimal schedules for the reference-point model. On the 24-h Starlink trace,
BL-BnB improves mean per-beam capacity over nearest-satellite selection by
27.2\% with eight beams and 9.8\% with 49 beams, primarily benefiting
weaker links. Increasing the number of overlapping beams raises total
capacity sublinearly, while performance varies little across the studied
cell radii. For selecting eight of 28 candidates, BL-BnB reduces mean
runtime from 529.48 to 1.98 ms, although worst-case complexity remains
exponential. These results support jointly considering link quality and
interference for satellite spatial reuse.

\begingroup
\microtypesetup{activate=true}
\def\UrlBreaks{\do\/\do\-\do\?\do\&\do\=}
\def\BIBdecl{%
    \let\paperoriginalbibitem\bibitem
    \renewcommand{\bibitem}[1]{\paperoriginalbibitem{##1}\looseness=-1\relax}}
\bibliographystyle{IEEEtran}
\bibliography{references}
\endgroup

\end{document}